# Forward Neutron Production at the Fermilab Main Injector

T. S. Nigmanov [k], D. Rajaram [k], M.J. Longo [k*], U. Akgun [j], G. Aydin [j], W. Baker [c], P.D. Barnes, Jr. [g], T. Bergfeld [l], A. Bujak [h], D. Carey [c], E.C. Dukes [m], F. Duru [j], G.J. Feldman [d], A. Godley [l], E. Gülmez [j,o], Y.O. Günaydin [j,p], N. Graf [f], H.R. Gustafson [k], L. Gutay [h], E. Hartouni [g], P. Hanlet [e], M. Heffner [g], C. Johnstone [c], D. M. Kaplan [e], O. Kamaev [e], J. Klay [g], M. Kostin [c], D. Lange [g], A. Lebedev [d], L.C. Lu [m], C. Materniak [m], M.D. Messier [f], H. Meyer [n], D.E. Miller [h], S.R. Mishra [l], K. S. Nelson [m], A. Norman [m,c], Y. Onel [j], J.M. Paley [f,q], H.K. Park [k,r], A. Penzo [j], R.J. Peterson [i], R. Raja [c], C. Rosenfeld [l], H. A. Rubin [e], S. Seun [d], N. Solomey [n], R. Soltz [g], E. Swallow [b], Y. Torun [e], K. Wilson [l], D. Wright [g], K. Wu [l]

*[a] Brookhaven National Laboratory, Upton, NY 11973*
*[b] Elmhurst College, Elmhurst, IL 60126*
*[c] Fermi National Accelerator Laboratory, Batavia, IL 60510*
*[d] Harvard University, Cambridge, MA 02138*
*[e] Illinois Institute of Technology, Chicago, IL 60616*
*[f] Indiana University, Bloomington, IN 47403*
*[g] Lawrence Livermore National Laboratory, Livermore, CA 94550*
*[h] Purdue University, West Lafayette, IN 47907*
*[i] University of Colorado, Boulder, CO 80309*
*[j] University of Iowa, Iowa City, IA 52242*
*[k] University of Michigan, Ann Arbor, MI 48109*
*[l] University of South Carolina, Columbia, SC 29208*
*[m] University of Virginia, Charlottesville, VA 22904*
*[n] Wichita State University, Wichita, KS 67260*
*[o] Also at Bogazici University, Istanbul, Turkey*
*[p] Now at Kahramanmaras Sutcu Imam University, Kahramanmaras, Turkey*
*[q] Now at Argonne National Laboratory, Argonne, IL 60439*
*[r] Now at Kyungpook National University, Daegu, 702-701, Korea*

** Corresponding author. Tel. 734 7644445; fax: 734 7639694: e-mail: mlongo@umich.edu*

**Abstract.** We have measured cross sections for forward neutron production from a variety of targets using proton beams from the Fermilab Main Injector. Measurements were performed for proton beam momenta of 58 GeV/*c*, 84 GeV/*c*, and 120 GeV/*c*. The cross section dependence on the atomic weight (*A*) of the targets was found to vary as $A^{\alpha}$ where $\alpha$ is $0.46 \pm 0.06$ for a beam momentum of 58 GeV/*c* and $0.54 \pm 0.05$ for 120 GeV/*c*. The cross sections show reasonable agreement with FLUKA and DPMJET Monte Carlos. Comparisons have also been made with the LAQGSM Monte Carlo.

**Keywords:** MIPP; Calorimeters; Neutrons; Hadrons; Fermilab

**PACS: 13,60.Hb, 13.85.Ni, 13.60.Rj, 13.75,Cs, 25.40.Ep**

# I. INTRODUCTION

The MIPP (Main Injector Particle Production) experiment (FNAL E907) [1] acquired data in the Meson Center beam line at Fermilab. The primary purposes of the experiment were to investigate scaling laws in hadron fragmentation [2], to obtain hadron production data for the NuMI (Neutrinos at the Main Injector [3]) target to be used for calculating neutrino fluxes, and to obtain inclusive pion, neutron, and photon production data to facilitate proton radiography [4].

While there is considerable data available on inclusive charged particle production [5], there is little data on neutron production. In this article we present results for forward neutron production using proton beams of 58 GeV/c, 84 GeV/c, and 120 GeV/c on hydrogen, beryllium, carbon, bismuth, and uranium targets, and compare these data with predictions from Monte Carlo simulations.

# II. APPARATUS

A schematic of the MIPP spectrometers is shown in Fig. 1. The detector consisted of two large aperture magnetic spectrometers. The "Jolly Green Giant" (JGG) and "Rosie" magnets each had a $p_t$ kick $\cong 0.32$ GeV/c and were operated with opposite polarity, so that their deflections approximately canceled. The incident beam entered from the left of the figure and struck targets located near the upstream entrance of the JGG. The beam particles were identified using Cherenkov counters upstream of the target. The beam included protons (antiprotons), kaons, and pions. Only data for incident protons are reported here.

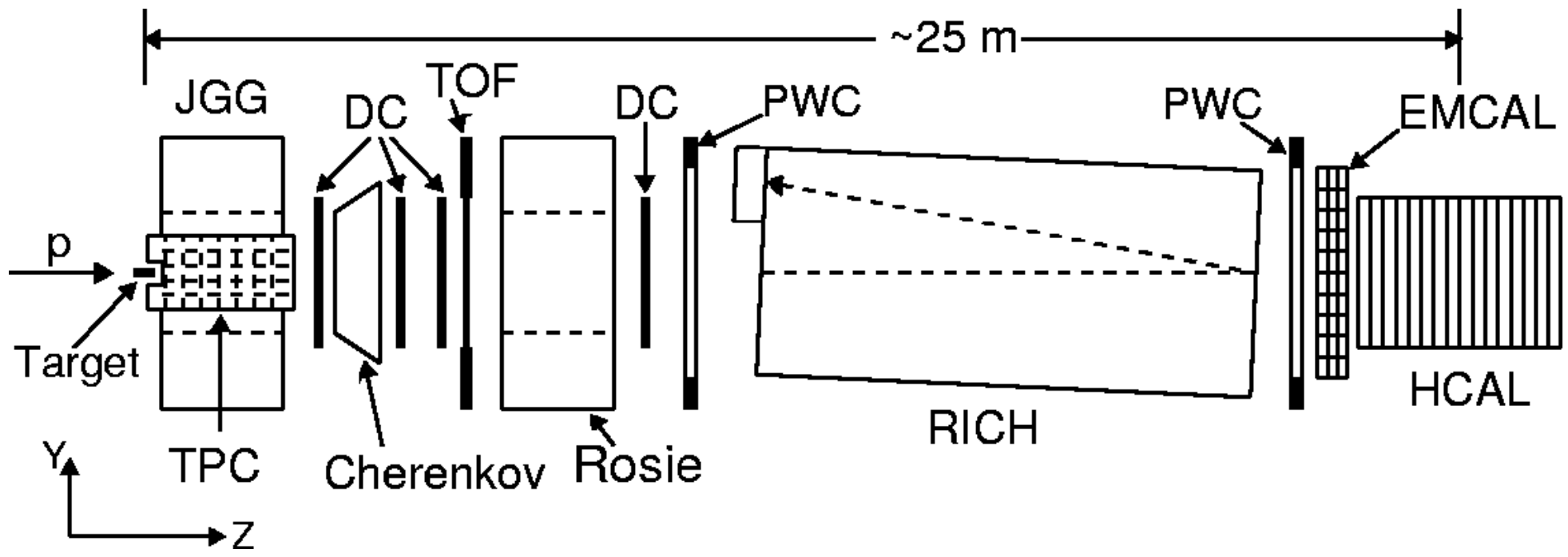

**FIGURE 1.** Experimental layout schematic.

The trajectories and momenta of the secondary charged particles were measured from hits in the time projection chamber (TPC), situated inside the JGG, and hits in the downstream drift chambers (DCs) and proportional wire chambers (PWCs). The track reconstruction was done in stages. Helical TPC track segments were first formed, followed by the formation of track segments using hits from the DCs and PWCs. Next, the TPC track segments were refit using the field of the JGG magnet, and these were then matched to the DC and PWC track segments that were fit using the field of the Rosie magnet. A fit of the trajectory was performed with all hit information included from the matched track segments. Primary and secondary vertices were then found by grouping secondary tracks that fall within some distance of closest approach. Finally, a vertex-constrained fit was performed using all tracks associated to each vertex. This fit simultaneously determined the vertex position and refit the particle trajectories such that all associated tracks originate from that vertex. The TPC provided charged particle identification (PID) in the low energy region (< 1 GeV) by means of ionization (dE/dx); the time-of-flight hodoscope (TOF) and Cherenkov detector provided PID in the intermediate region (1 – 17 GeV/c); and the ring-imaging Cherenkov counter (RICH) provided PID for high energy tracks (>17 GeV/c). Neutrons were identified and their energies measured through their interactions in the electromagnetic shower detector and hadron calorimeter.

The solid targets were placed in the beam ~7 cm upstream (the liquid hydrogen target was ~12 cm upstream) of the entrance to the TPC. The hydrogen target was 14 cm long and 3.8 cm in diameter. The solid targets ranged in thickness from 0.17 cm to 1.0 cm; all had a radius of 2.54 cm. Immediately following the target was a 0.32 cm thick, 7.6 cm x 5.1 cm scintillation counter (SCINT) which was used to form the interaction trigger. The threshold was based on the pulse height response of the counter. It had an efficiency of ~95% for events with 3 tracks and (due to the Landau tail in the scintillation response) ~1% efficiency for single track events. Most of the data were taken with this trigger requirement. In addition, prescaled "beam triggers" were collected to count the incident beam flux and also calibrate the efficiency of the SCINT trigger. For this analysis only triggers consistent with an incoming proton were selected. The physical properties of the targets are listed in Table 1.

| **TARGET** | **A** | **d** (cm) | **AD** ($g/cm^2$) | **IL** ($\lambda_I$) | $\mathbf{n_t}$ ($10^{23}$ $cm^{-2}$) |
|---|---|---|---|---|---|
| $H_2$ (liquid) | 1.008 | 14.0 | 0.991 | 0.015 | 5.922 |
| Beryllium | 9.012 | 0.399 | 0.71 | 0.0094 | 0.4744 |
| Carbon | 12.011 | 1.003 | 1.677 | 0.0194 | 0.8408 |
| Bismuth | 208.98 | 0.173 | 1.69 | 0.0087 | 0.0487 |
| Uranium | 238.02 | 0.1 | 1.875 | 0.0110 | 0.0474 |

**TABLE 1.** The targets and their properties, where d is thickness in cm, AD is the areal density in $gm/cm^2$, IL is the number of interaction lengths and $n_t$ is the number of nuclei per $cm^2$. $n_t$ was calculated as: $n_t = N_A \times density \times thickness / A$ where $N_A$ is Avogadro's number and $A$ is the atomic weight of the target material.

The electromagnetic and hadron calorimeters allowed us to measure the production of forward-going long-lived neutral particles – photons and neutrons – that are not observable in the upstream detectors. The electromagnetic calorimeter was built for the MIPP experiment, while the hadron calorimeter was originally built for the HyperCP (E871) experiment [6]. A schematic of the two calorimeters is shown in Fig. 2. The electromagnetic calorimeter (EMCAL) consisted of 10 layers of 5.08 mm thick lead interspersed with planes of gas proportional chambers. The proportional chambers were made from 1.5 m long aluminum extrusions. There were 64 anode wires with 2.54 cm spacing in each plane. Alternate planes had wires oriented horizontally and vertically. The chambers used a gas mixture of 76.5% Argon, 8.5% Methane, and 15% $CF_4$. The EMCAL active volume was 1.6 m wide, 1.5 m high, and 0.3 m in the beam direction. Its total thickness was ~10 radiation lengths. The EMCAL pulse height readout system consisted of 640 amplifier channels with multiplexed 12-bit ADCs [7]. The hadron calorimeter (HCAL) was composed of 64 layers of 24.1 mm iron plates interspersed with 5 mm thick scintillators as the active medium [6]. The total thickness of the HCAL was 9.6 interaction lengths (88.5 radiation lengths). Its active area is 0.99 m wide, 0.98 m high, and 2.4 m in the beam direction. For readout purposes the HCAL was subdivided into four longitudinal and two lateral sections, for a total of 8 cells that were read out with wavelength shifting fibers spaced 30 mm apart. Fibers from each cell were bundled into a single 2-inch Hamamatsu R329-02 photomultiplier tube with extended green sensitivity. The pulse heights were flash digitized in custom built CAMAC 14-bit ADC modules [6] with a 75 fC least count.

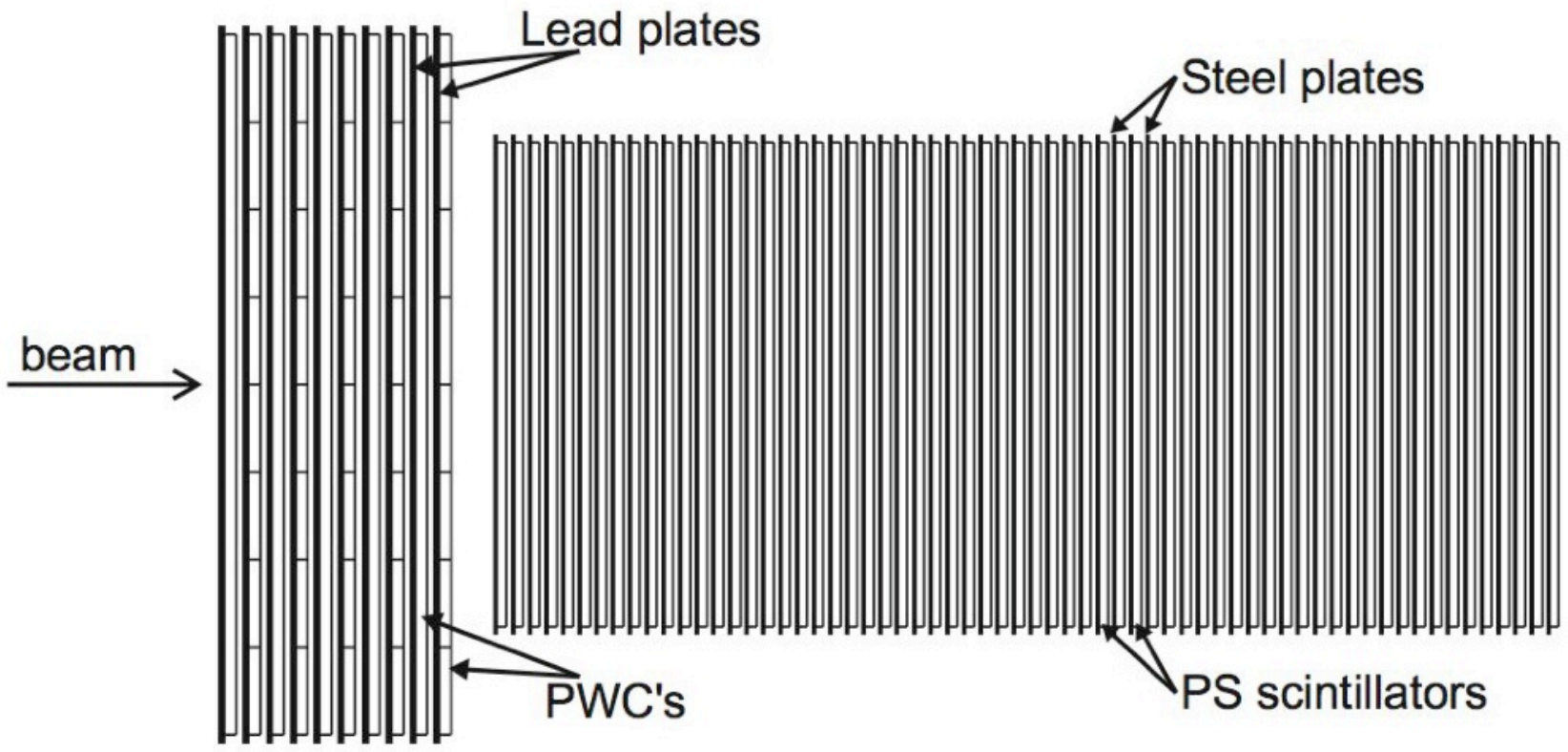

**FIGURE 2.** Schematic of calorimeters (not to scale).

The calorimeters were calibrated with incident hadron and electron beams of various momenta and are described in [8]. Table 2 lists the energy resolution of the calorimeters, and Fig. 3 shows the resolution as a function of the proton beam momentum.

| Particle | p (GeV/c) | σ/E (%) |
|---|---|---|
| $e$ | 18.5 | 6.2±0.3 |
| $p$ | 20 | 13.8±1.4 |
| $p$ | 35 | 10.7±0.9 |
| $\pi$ | 58 | 7.6±0.3 |
| $K$ | 58 | 7.6±0.3 |
| $p$ | 58 | 7.6±0.3 |
| $p$ | 84 | 6.7±0.2 |
| $p$ | 120 | 5.9±0.4 |

**TABLE 2.** The energy resolution of the calorimeters for various particle species and momenta.

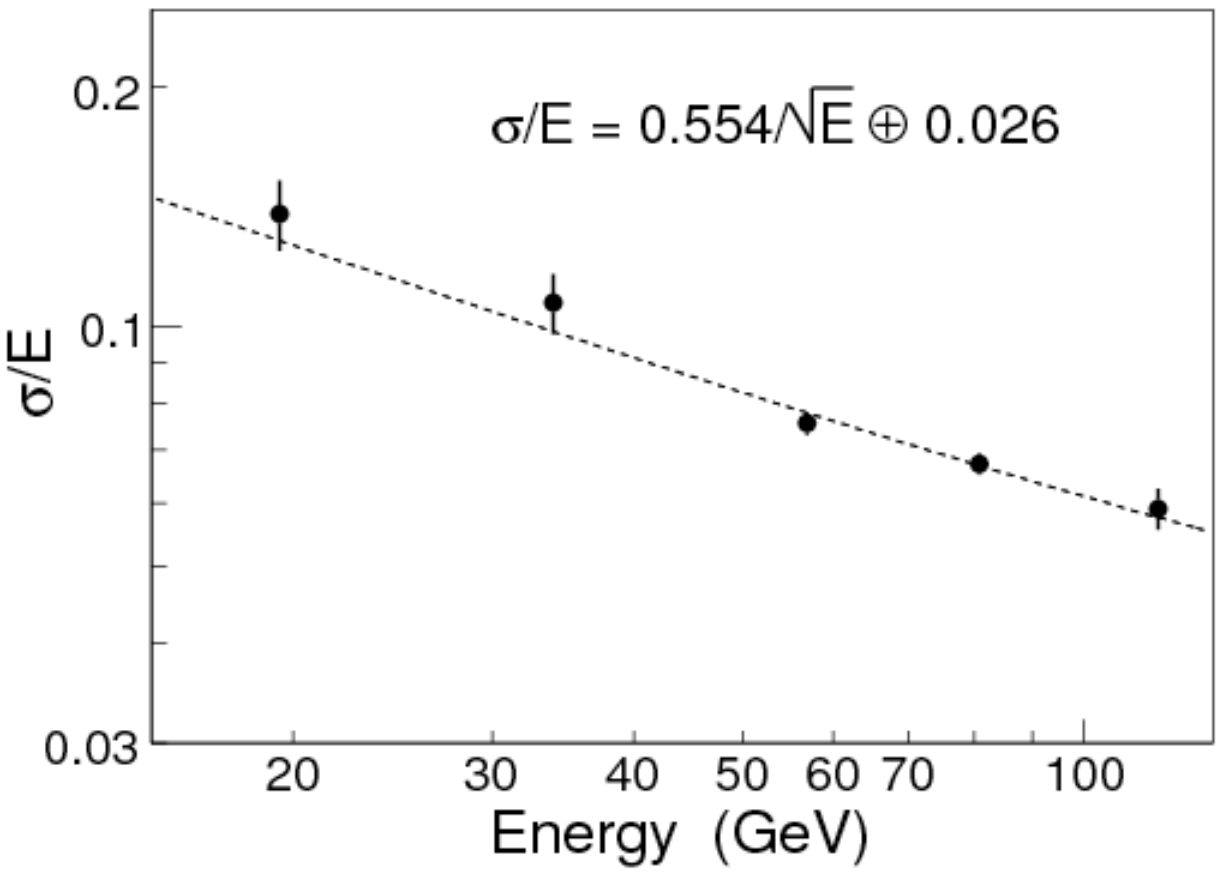

**FIGURE 3.** The fractional energy resolution of the calorimeters as a function of the proton beam energy. The curve represents the fit to $\sigma / E = a / \sqrt{E} \oplus b$, where E is in GeV. "$\oplus$" indicates addition in quadrature.

# III. MONTE CARLO SIMULATIONS

The main purposes of the Monte Carlo simulations were to determine the geometric acceptance of the calorimeters, the SCINT trigger efficiencies, and the backgrounds and selection efficiencies. These are only weakly model dependent. An important goal of the experiment was to compare the neutron production data with the predictions of current Monte Carlo simulations of hadron production.

The Monte Carlo was based on FLUKA2006 [9, 10] for the production of the secondaries and an implementation of GEANT 3.21 [11] for their propagation. (For the hydrogen target, DPMJET [12,13] was used as the event generator.) Monte Carlo events were run through the same analysis as the data and the same event selection criteria were applied. The spatial and momentum distribution of the incident beam, the energy resolution of the calorimeters, the momentum resolution for charged particles, and the spatial resolution of the TPC and wire chambers were all simulated.

In Section VII we compare the measured neutron distributions with those predicted by FLUKA(DPMJET) and with the predictions of the LAQGSM model [14]. Note that the LAQGSM modeling has been performed with the MARS15 code [14] with the LAQGSM event generator [15,16,17] using the 2010 version of LAQGSM.

# IV. EVENT AND NEUTRON SELECTION

In this section we describe first the incident beam selection and then the neutron event selection. The incident proton beam flux, $N_{beam}$, was determined by counting the number of unbiased proton beam trigger events and applying the run-dependent prescale factor that was set during data-taking. There was a hardware requirement that the SCINT trigger fired, indicating an interaction in the target that produced ionization equivalent to 3 or more charged particles. This removed most of the events with fewer than 3 forward-going charged particles.

For the overall event selection, we imposed the following conditions in the analysis:

1) To select clean events, there must be no more than 30 reconstructed secondary charged tracks in an event.

2) There must be only one beam track incident on the target so that the initial state is well determined.

3) The transverse positions of the beam track must be consistent with the dimensions of the target. The transverse position distribution of the incident beam is shown in Figure 4.

To be considered for the neutron analysis an event had to pass the following further conditions:

4) Events were required to have a primary vertex no more than 4 cm upstream and 6 cm downstream of the target along the beam direction. For the 14 cm long liquid hydrogen target

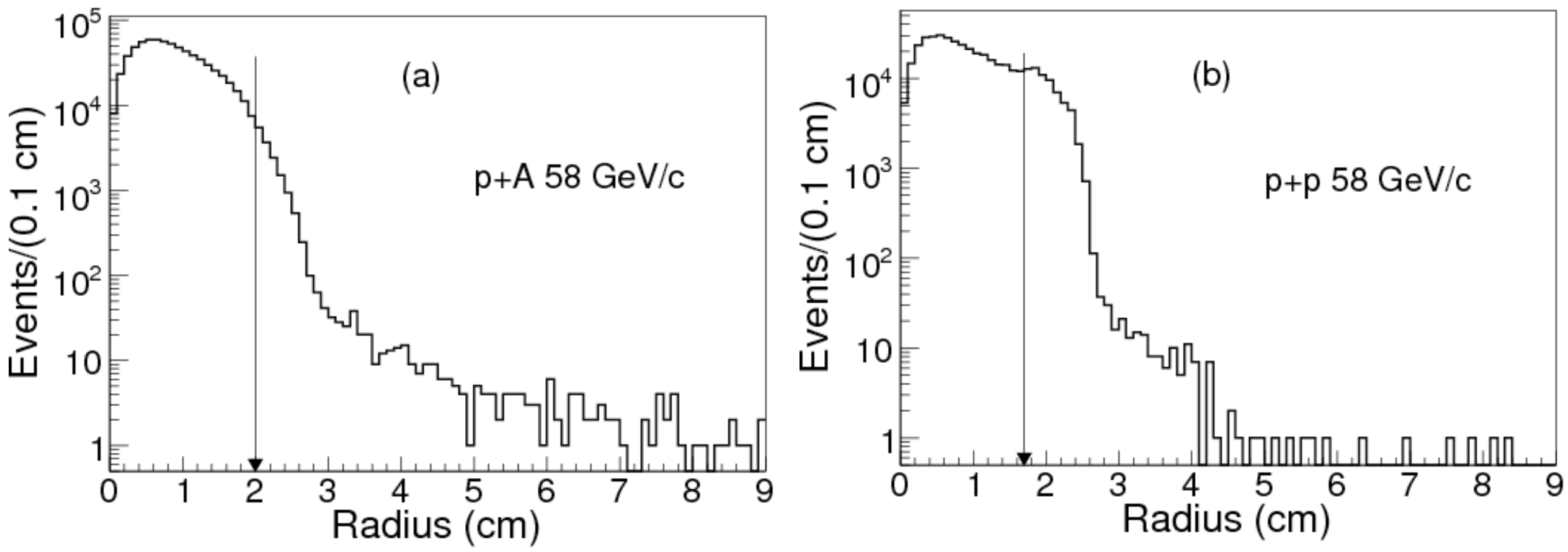

**FIGURE 4.** Radial positions of 58 GeV/c incident protons on solid target (left) and hydrogen target (right), based on beam triggers. The arrows indicate the location of the selection cuts. The profile for the hydrogen target is broader because of additional material just upstream of the target.

we required that the vertex be within 15 cm of the target center. Typical longitudinal positions of the reconstructed vertex are shown in Fig. 5. The longitudinal requirements encompassed the SCINT counter. Interactions from the SCINT were then subtracted by analyzing the “target-out” data as described later. Appropriate cuts on the transverse position were also applied

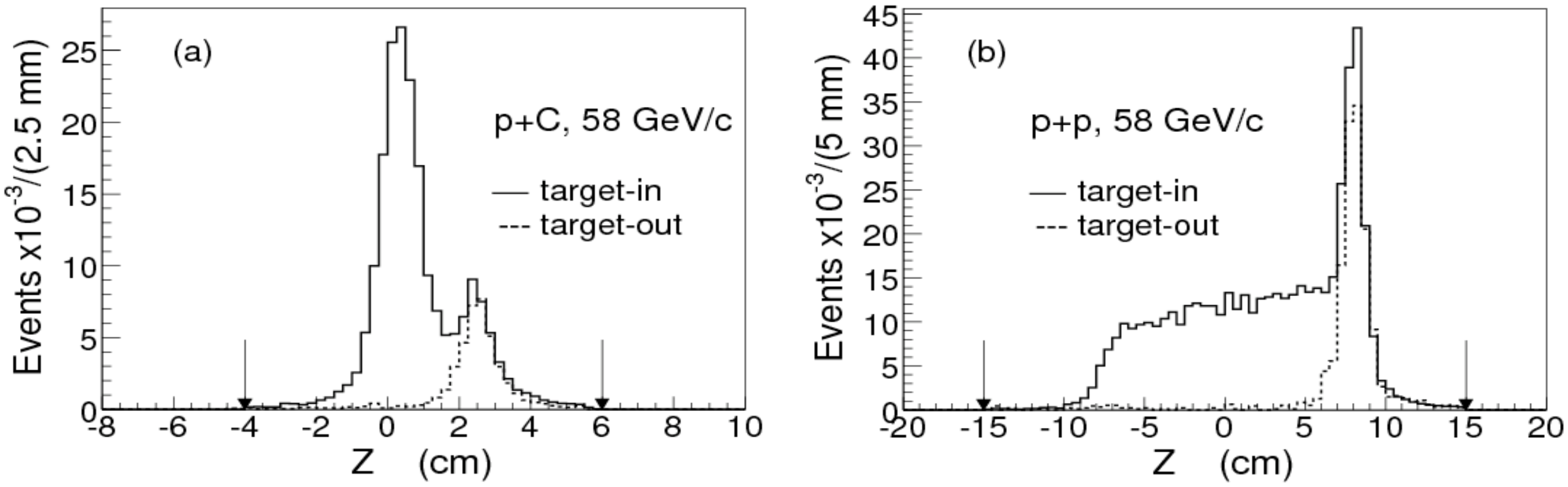

**FIGURE 5.** Longitudinal positions of reconstructed vertices from carbon target (left) and hydrogen target (right).The arrows indicate the location of the selection cuts. The carbon target was 1 cm thick and 5.08 cm in diameter, while the hydrogen target was 14 cm thick and 3.8 cm in diameter.

5) To reduce contamination of the sample by beam protons that appeared to show small deflections in the target ("straight-throughs"), the difference in transverse momentum between the incident beam track and the sum of that of the outgoing tracks ($\Delta p_T$ ) was required to be at least 150 MeV/c. Figure 6 shows a typical distribution of the “missing” transverse momentum.

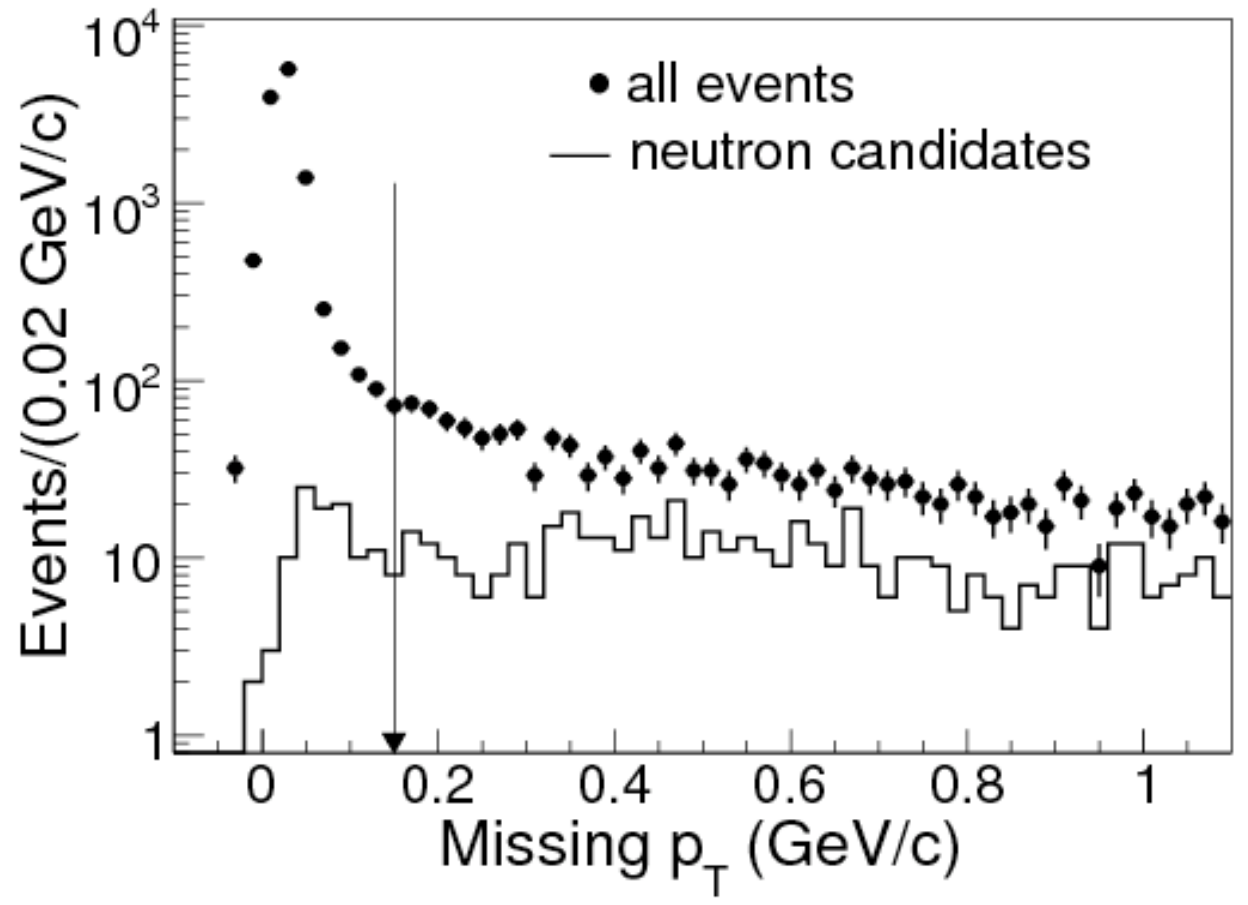

**FIGURE 6.** The difference in transverse momentum between the incident beam particle and the outgoing secondary tracks. The arrow indicates the location of the cut. The peak at 0 represents straight-through beam tracks that were rejected by the cut.

6) Despite the $\Delta p_T$ requirement above, there was still some contamination of straight-through tracks due the Landau tails of the SCINT pulse height response. In order to eliminate remaining uninteracted beam particles, the event was rejected if any charged track projected into the calorimeter's fiducial area and had momentum > 0.7 $p_{beam}$.

As a cross-check, the $N_{beam}$ calculated above was compared with the counts from scalers accumulated during data-taking and the results were found to be consistent to within 7%. A conservative systematic uncertainty of 10 % was then assigned to the incident beam flux. As an overall cross-check, we measured the total inelastic cross sections for p+p interactions at 58 GeV/c and 84 GeV/c. The results from our data listed in Table 3 were found to be consistent within the beam flux uncertainty with known values from [18].

| | Inelastic Cross Section (mb) | |
|---|---|---|
| | 58 GeV/c | 84 GeV/c |
| MIPP | $29.2 \pm 2.9$ | $33.7 \pm 3.4$ |
| PDG | $31.0 \pm 0.3$ | $31.4 \pm 1.0$ |
| DPMJET | 30.6 | 30.9 |

**TABLE 3.** Our measurement of total inelastic cross sections for p+p interactions at 58 GeV/c and 84 GeV/c compared with PDG values [18] and predictions from DPMJET Monte Carlo.

Neutron candidates were identified by measuring the energy deposited in the calorimeters and then subtracting from this the energies of charged tracks within the geometric acceptance of the calorimeter. It was further required that the neutron energy thus measured should exceed ~20% of the beam energy. Thus, the minimum neutron energies were 12, 18, and 20 GeV for beam momenta of 58, 84, and 120 GeV/c respectively. To calibrate the calorimeter response, the average energy loss of neutrons in the EMCAL due to interactions with lead nuclei was estimated using the average energy loss of protons in the EMCAL. (The energy loss of protons in the EMCAL due to ionization is <0.08 GeV.)

Figure 7 shows the energy deposition of protons in the EMCAL as a function of beam momentum. As can be seen, at the momenta of interest less than 10% of the energy is typically lost in the EMCAL. The final measured neutron energy then is $E_{emcal} + E_{hcal} - \sum E_{trk,cal}$ , where $E_{hcal}$ and $E_{emcal}$ are the energies deposited in the HCAL and EMCAL, and $\sum E_{trk,cal}$ is the summed energy of all charged tracks heading into the HCAL. Note that since the poor transverse segmentation in the HCAL did not allow us to localize the neutron interaction vertex, all measurements reported here are integrated over the neutron transverse momentum, $p_T$.

Target-out subtractions were made to correct for scattering from material near the target, in particular, the trigger scintillator and, for the hydrogen target, the target flask and vacuum windows.

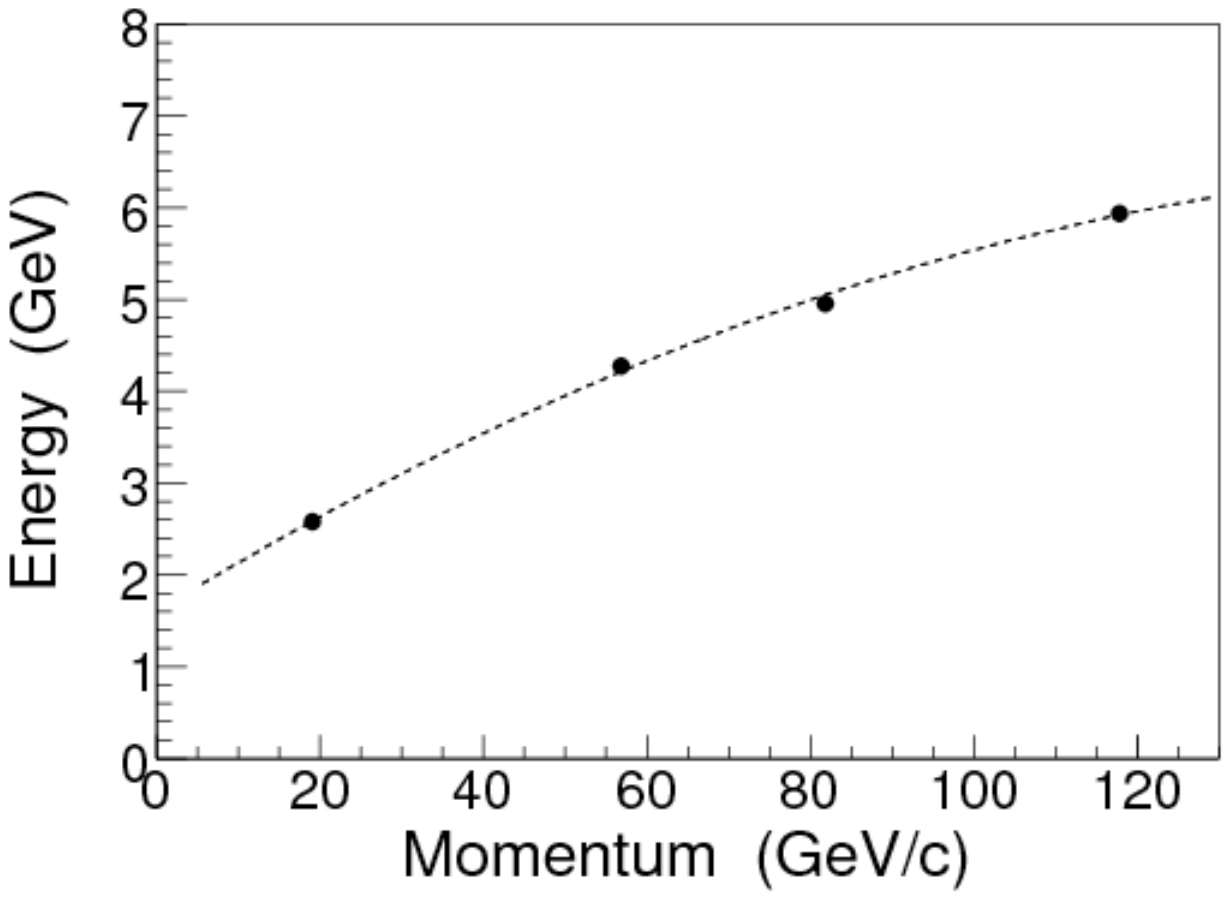

**FIGURE 7.** Average energy loss of protons in the EMCAL. The energy loss of neutrons in the EMCAL was assumed to be similar to that of protons.

Figure 8 shows examples of the raw target-in and target-out neutron spectra before correcting for background, geometric acceptance, and triggering efficiency. The target-out contributions were relatively large because the targets used were thin (~0.01 interaction lengths, see Table 1) compared to the amount of material surrounding the targets. The target-out data were normalized to correspond to the number of incident protons in the target-in data.

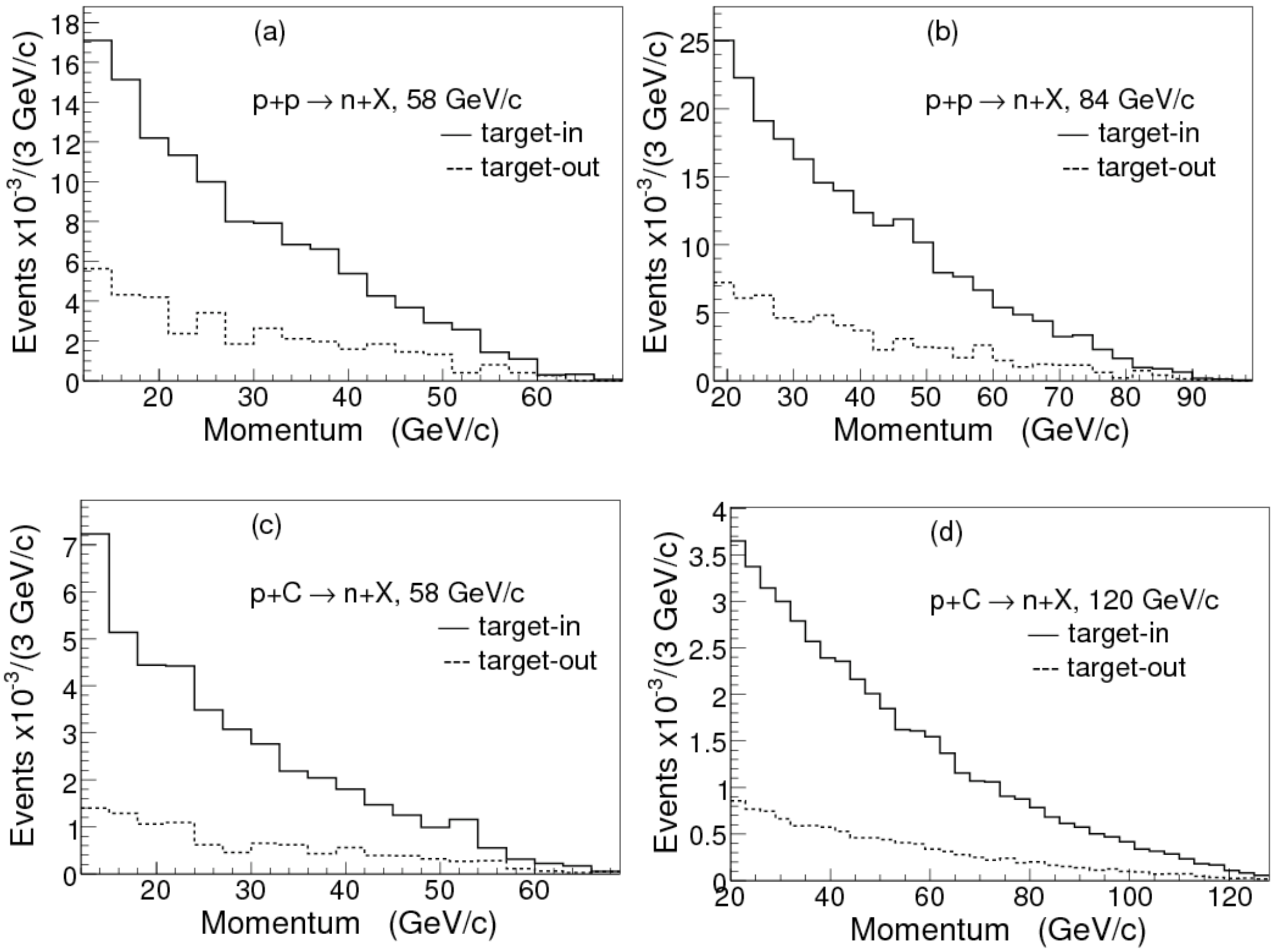

**FIGURE 8.** Raw momentum spectra of neutron candidates before acceptance, trigger, and background corrections. Plots on the top are from the hydrogen target and those on the bottom from the carbon target.

# V. CORRECTIONS

In addition to the target-out corrections described above, the Monte Carlo was used to correct for the following:

1) The inefficiency for triggering on neutron events.

2) Efficiency of the neutron selection requirements (vertex position and $\Delta p_T$ cuts discussed above).

3) Neutron losses due to interactions with material in the spectrometer.

4) The contamination of $K_L^0$, secondary neutrons, and photons.

We studied the SCINT trigger efficiency using unbiased beam triggers and also with Monte Carlo. Figure 9 shows the typical dependence of the trigger efficiency on momentum. The efficiency is relatively high at low neutron momentum and drops at higher momentum due to the low charged particle multiplicity associated with high momentum neutrons.

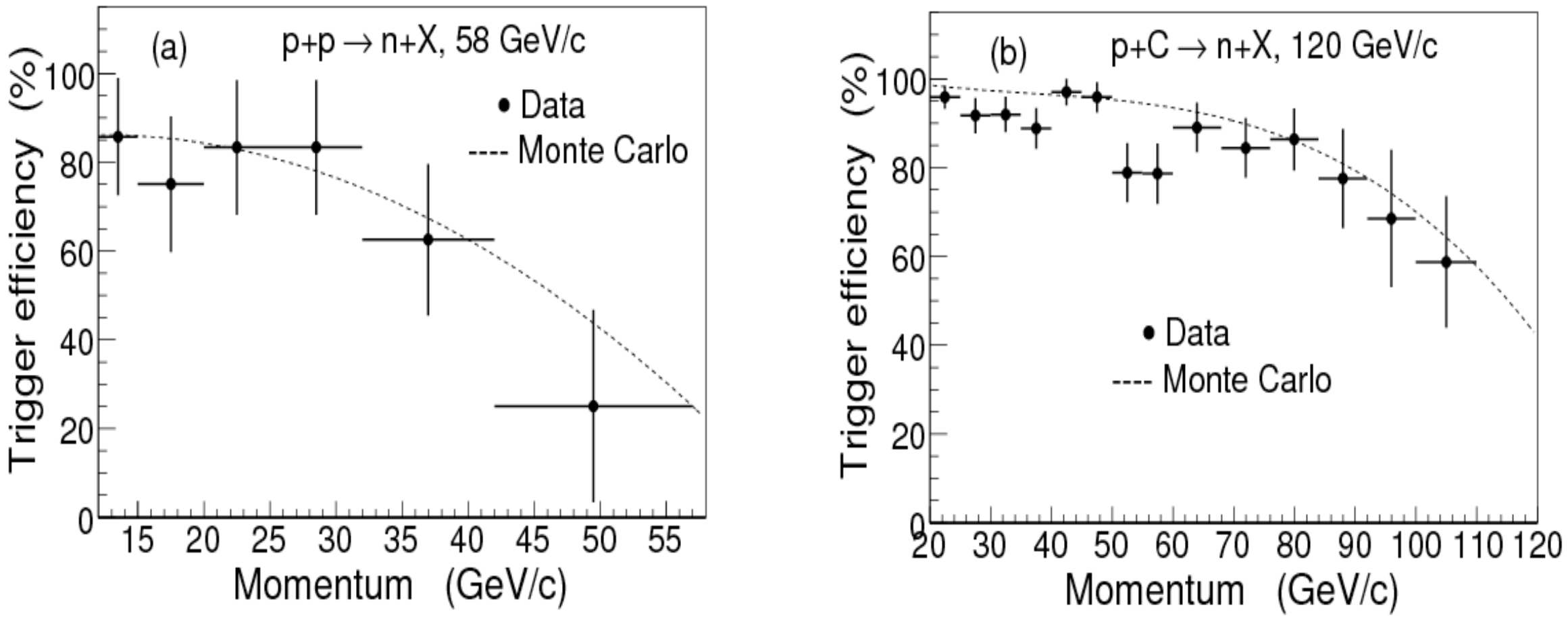

**FIGURE 9.** The SCINT trigger efficiency as a function of neutron momentum for data (solid points) and Monte Carlo (dashed curve). Note that the horizontal axis minimum is at the neutron selection threshold and not at 0.

In Figure 10 we show the dependence of the neutron selection efficiency on momentum. The efficiency appears to be almost independent of momentum, as expected.

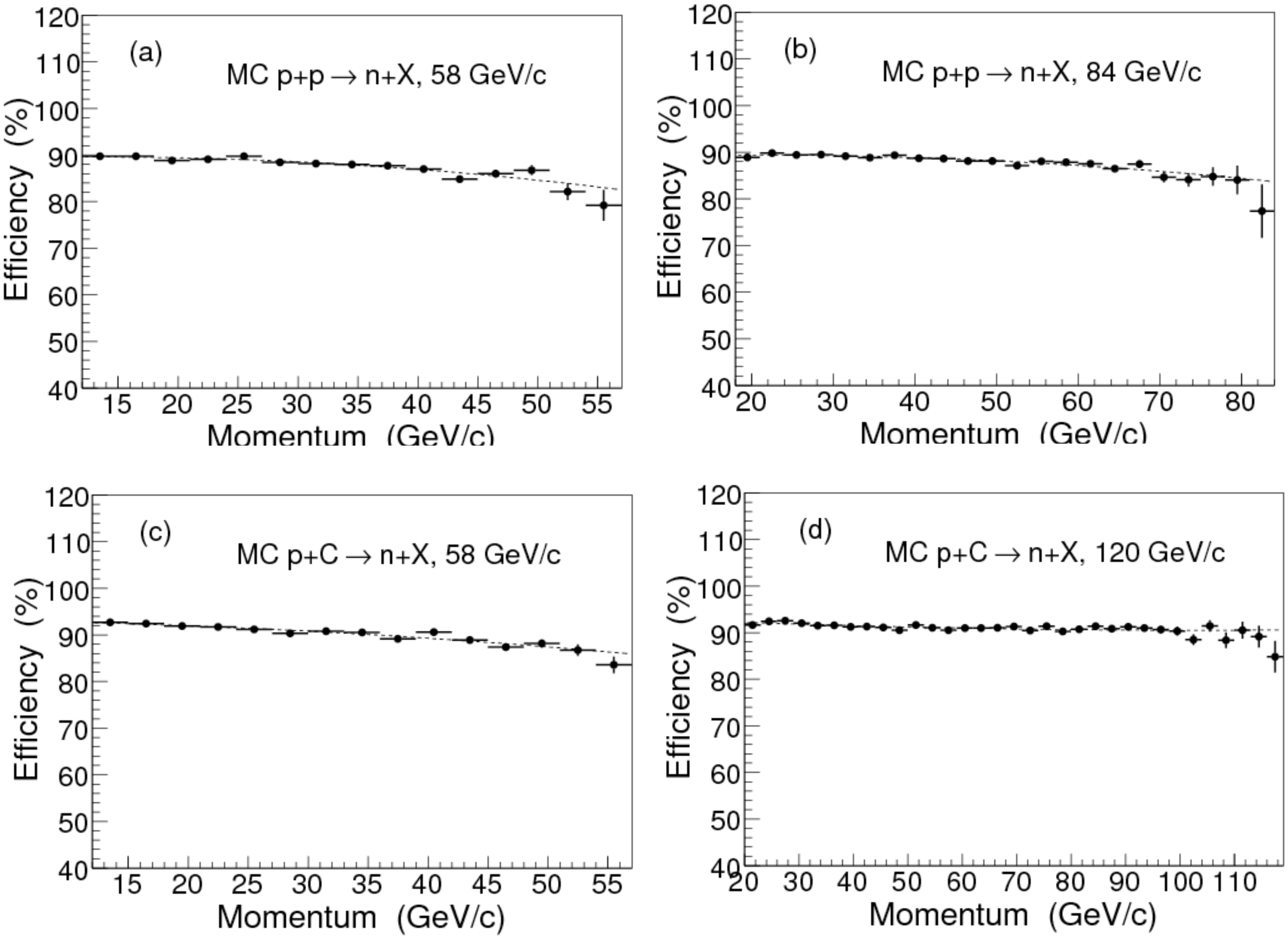

**FIGURE 10.** Dependence of the neutron selection efficiency on neutron momentum. This includes the $\Delta p_T$ cut and requirements on the position of the production vertex, and does not include the geometric acceptance (discussed below) or trigger efficiency (already shown above in Figure 9).

Neutrons produced in the target could interact with the spectrometer material and not be detected in the calorimeter. Monte Carlo estimates show the following loss rates: ~7% in the TOF, ~3% in the walls of the RICH, and ~2% in the rest of the spectrometer.

Figure 11 shows the momentum-dependent contributions of $K_L^0$, secondary neutrons, and photons as a fraction of the predicted neutron cross section for the hydrogen and carbon targets. $K_L^0$ are mainly produced in the target. Secondary neutrons are neutrons produced from charged particles interacting in the TOF counters, walls of the RICH, and other spectrometer material. The contribution of photons is the leakage of gammas from electromagnetic showers that were not fully contained in the EMCAL and entered the HCAL. The excess of secondary neutrons at high momentum is due to the p-n elastic charge exchange process simulated by GEANT. The background is comparable to the signal in the lowest momentum bin and is ~10% at higher momenta.

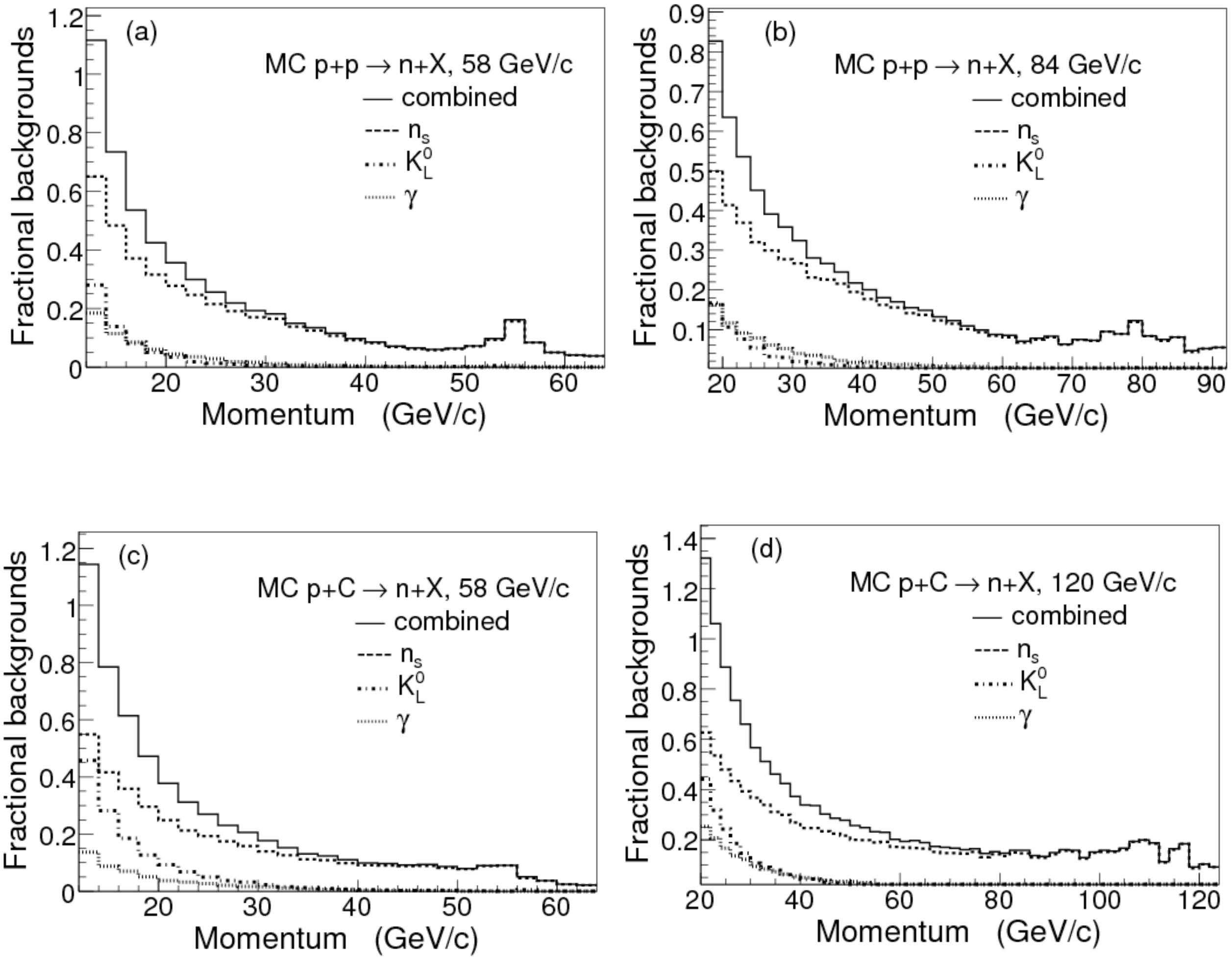

**FIGURE 11.** Typical momentum dependence of backgrounds predicted by the Monte Carlo as a fraction of the predicted neutron production cross section.

As a cross-check on the backgrounds in the Monte Carlo, we compared the predicted $K^0$ and $\pi^0$ multiplicities and cross sections with known measurements [19], and they were found to be consistent as shown in Table 4.

| | Average Multiplicity | | Cross section (mb) | |
|---|---|---|---|---|
| | Data | MC | Data | MC |
| $\pi^0$ | $2.68 \pm 0.3$ | 2.91 | $83.0 \pm 4.0$ | 89.6 |
| $K_s^0$ | $0.14 \pm 0.02$ | 0.18 | $4.3 \pm 1.0$ | 5.5 |
| $K_L^0$ | n/a | 0.18 | $4.3 \pm 1.0$ | 5.5 |

**TABLE 4.** Comparison of average multiplicities and production cross sections of neutrals for data [19] and Monte Carlo, from p+p interactions at 84 GeV/c.

As an overall cross-check to ensure that the Monte Carlo spectra accurately reproduced all effects in the data, we compared our data with the fully reconstructed Monte Carlo. Figure 12 presents the raw neutron spectra without any corrections applied. The Monte Carlo spectra show the reconstructed hadron calorimeter response where neutrons were identified and measured the same way as with data. It should be noted that the efficiencies and backgrounds discussed above have been shown only for illustration. In practice for each target/momentum, all the effects are combined under a single momentum-dependent correction. To do this, we compared the Monte Carlo-generated "true" neutron spectrum with the neutron spectrum found after the Monte Carlo events were reconstructed in exactly the same way as was done with data. The "true" and reconstructed spectra are shown in Figure 13. The ratio of the two spectra gives the combined correction to account for the effects listed above. For each target/momentum, the ratio was fit and the resulting corrections were applied to the data.

It should be noted that the reconstructed spectrum comes from the entire Monte Carlo sample that includes all generated events. Events containing a high-energy neutron are only a small fraction of this sample (e.g., the cross section for neutron production from hydrogen is ~6 mb, compared with the total inelastic cross section of ~31 mb).

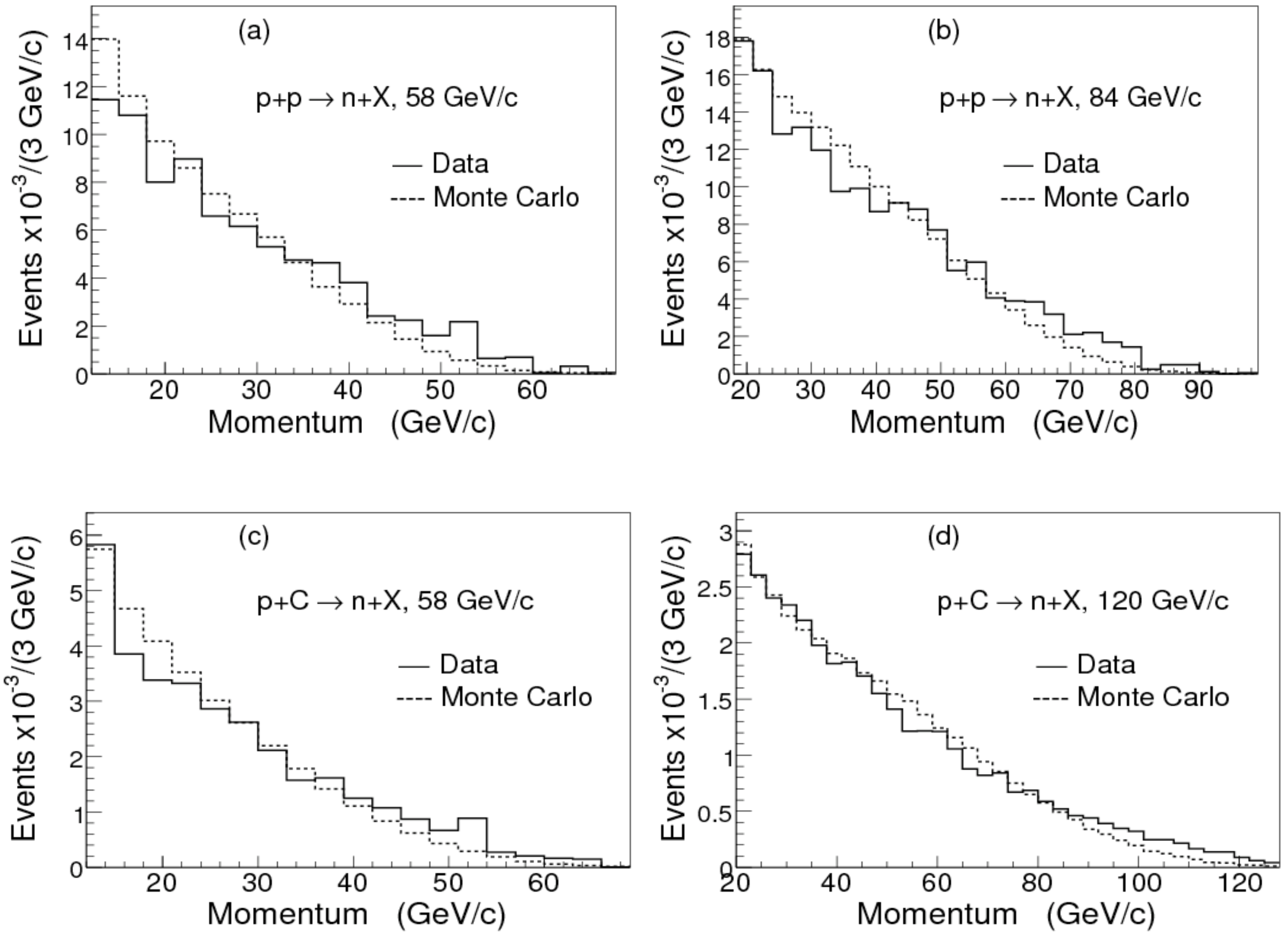

**FIGURE 12.** Comparison of data (solid) with fully reconstructed Monte Carlo (dashed) neutron candidate events. The Monte Carlo events have been normalized to have an equal number of total events and hence only the shapes are to be compared in this plot. The agreement in shapes indicates that all the reconstruction and resolution effects have been properly simulated in the Monte Carlo.

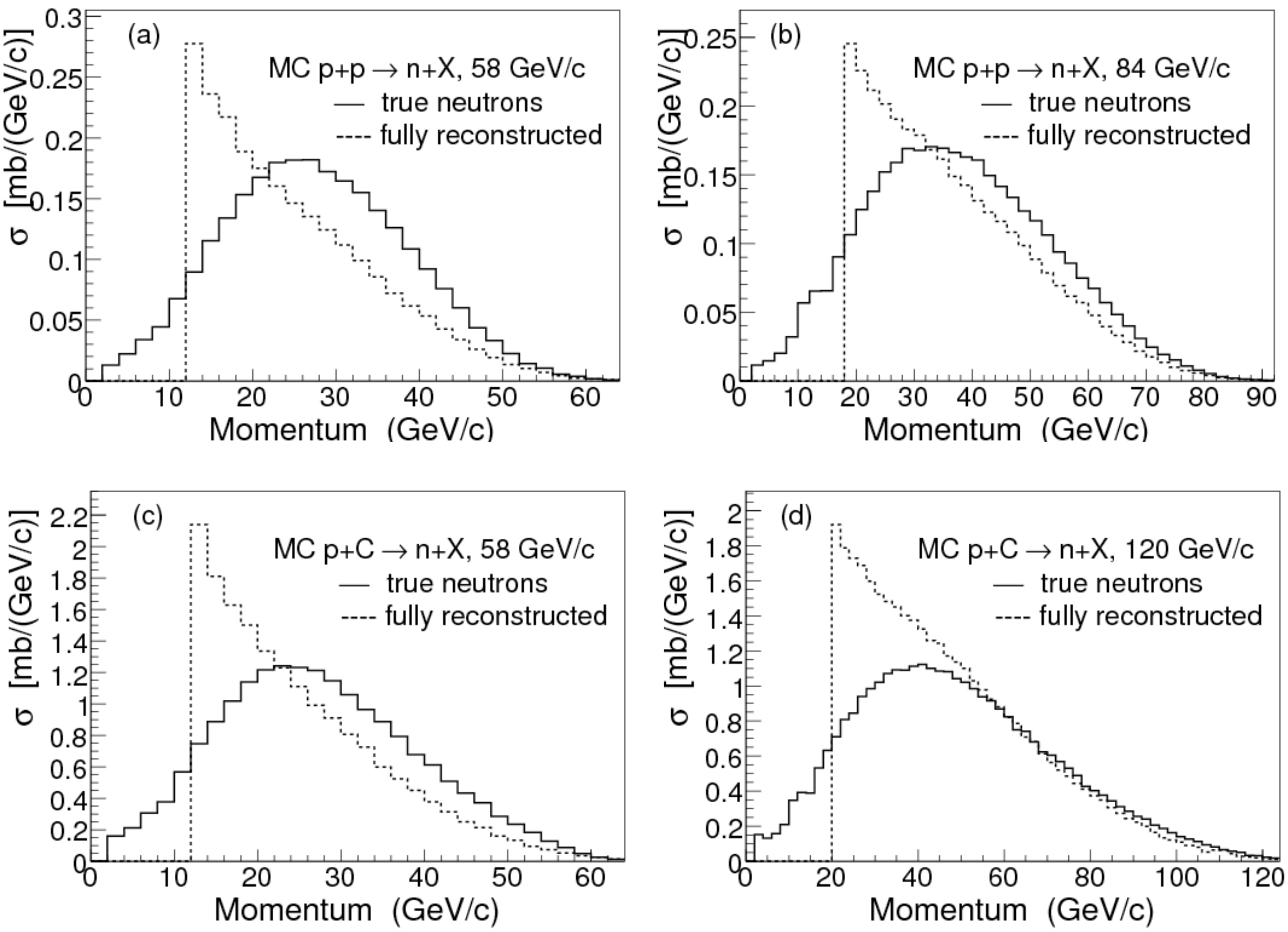

**FIGURE 13.** "True" neutron spectra generated by Monte Carlo (black) compared with reconstructed neutron candidates from the full Monte Carlo sample (dashed)

At neutron momenta lower than ~24 GeV/c for 58 GeV/c beam and neutron momenta less than 60 GeV/c for 120 GeV/c, the excess in the reconstructed spectrum is caused by backgrounds (see Fig. 11), whereas the deficit at higher momenta is due to triggering and selection efficiencies (see Figs. 9 and 10), and neutron losses due to interactions with spectrometer material.

The systematic uncertainty due to the correction was determined by varying the effect of the correction function by $\pm 30\%$ to be conservative. The neutron spectra from MIPP data after applying the combined correction are shown in Figure 14.

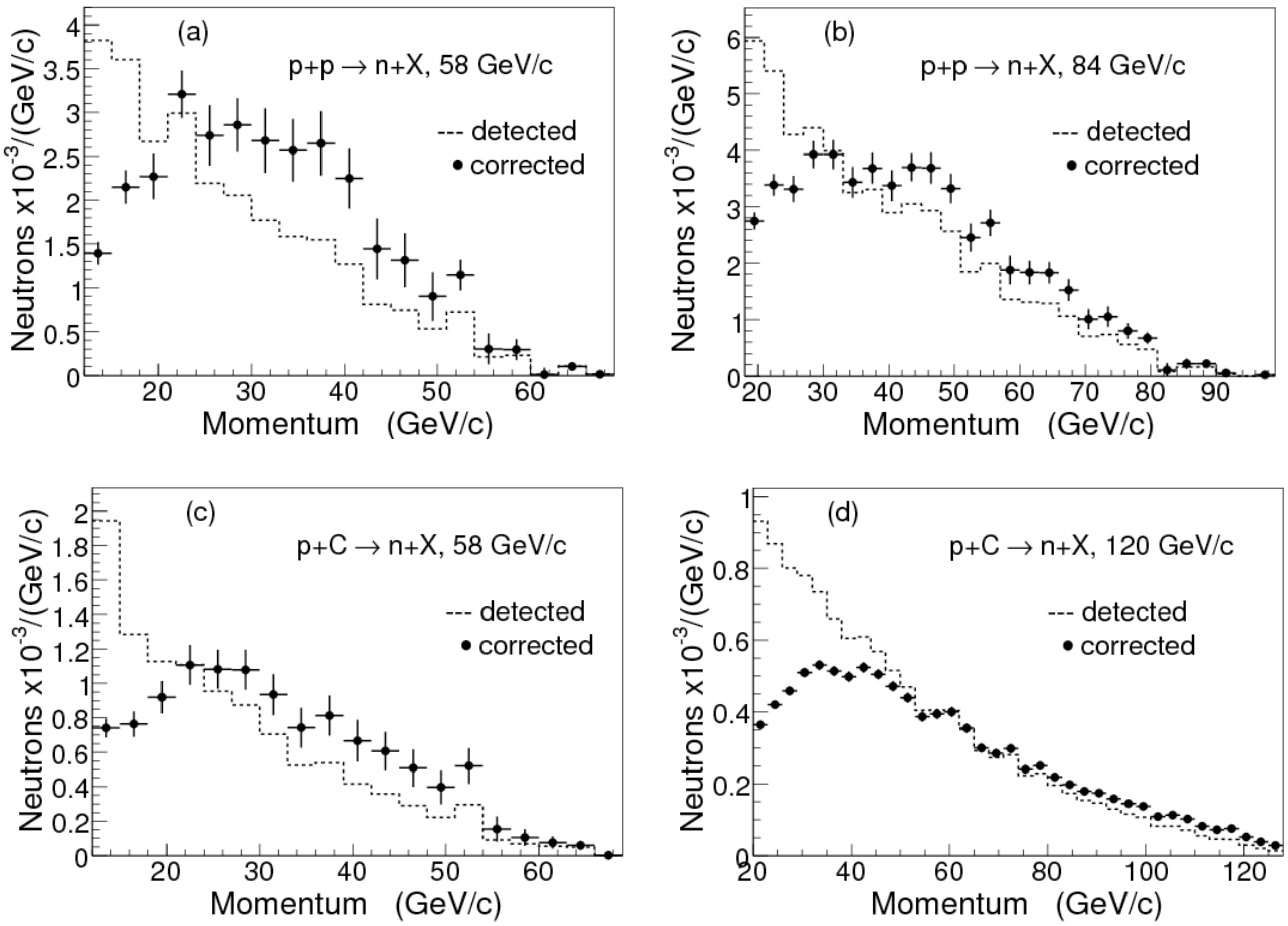

**FIGURE 14.** Neutron spectra from MIPP data after correction. The correction, as described earlier in the section, includes backgrounds, neutron losses and efficiencies for triggering and selection. The uncertainties are statistical.

# VI. ACCEPTANCE

In this section we discuss the geometric acceptance of the HCAL. The fiducial volume of the calorimeter for neutrons was defined as a circular region of radius 45 cm centered on the beam line at a *z* position midway through the hadron calorimeter. This subtended an angle of ~20.4 mrad from the target center.

The acceptance of the calorimeter for neutrons was determined using the FLUKA/DPMJET Monte Carlo that included the size of the incident beam and its momentum distribution. For each generated neutron having momentum greater than the threshold, we asked if it projected into the calorimeter's fiducial region. Then for each momentum bin, the fraction of neutrons that fell into the calorimeter gives the geometric acceptance for a given beam momentum. This fraction will depend on the transverse momentum distribution of neutrons generated in the MC, and thus is model dependent. For comparison, the acceptance was also determined using LAQGSM. Figure 15 shows the acceptance as a function of the neutron's momentum for the hydrogen and carbon targets for both event generators. At 58 GeV, LAQGSM gives a larger acceptance than both DPMJET and FLUKA, especially at higher neutron momenta. In contrast, at 120 GeV/c, LAQGSM agrees better with FLUKA at higher neutron momenta, but gives lower acceptance at lower neutron momenta. It should be noted that the comparison with LAQGSM was done only to illustrate the differences between different models. All acceptance corrections applied in this analysis were based on DPMJET or FLUKA. To determine the systematic uncertainty in the acceptance, we conservatively varied the acceptance by $\pm 30\%$ and found the corresponding fractional variation in the cross section.

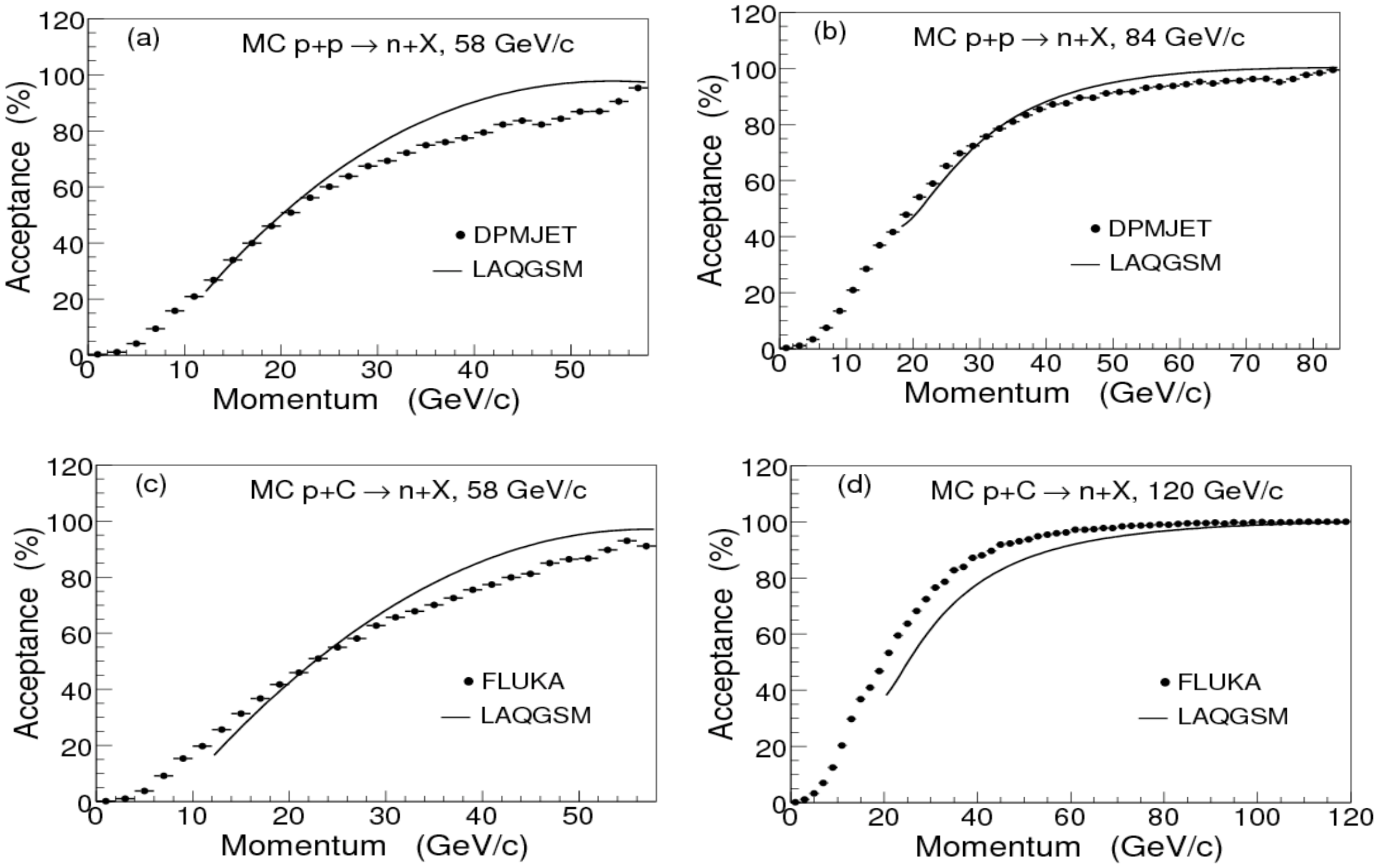

**FIGURE 15.** Geometric acceptance of the calorimeter for neutrons from hydrogen (top) and carbon (bottom) based on DPMJET/FLUKA and LAQGSM models.

## VII. RESULTS

The partial cross section $d\sigma / dp$ for producing a neutron within the fiducial volume of the hadron calorimeter (corresponding to an angular range of 20.4 mrad) with energy greater than the neutron threshold (of ~20% of the beam energy) is calculated as

$$\frac{d\sigma}{dp} = N_n \times \frac{1}{N_{beam}} \times \frac{1}{n_t} \times \frac{1}{b_s} \times 10^4, \quad \text{mb} / (\text{GeV} / c) \qquad (1)$$

where $N_n$ is the observed number of neutrons corrected for inefficiencies, losses and backgrounds as described in Section V; $N_{beam}$ is the corresponding number of incident protons in the beam; $n_t$ is the number of nuclei per cm$^2$ in the target (see Table 1), $b_s$ is width of the momentum bins, and the factor $10^4$ is to bring results to mb units.

Below we present our results in different ways:

1) Cross sections without correcting for the geometric acceptance of the calorimeter. This is done to make detailed comparisons with MC models and also to study the *A*-dependence,
2) Lorentz-invariant cross sections to study scaling,
3) Cross sections corrected for the geometric acceptance of the calorimeter,
4) Neutron density distributions in order to compare with results from NA49 [23].

In Figure 16 we show neutron cross sections vs. beam momentum for various targets and beam momenta. The error bars on the data points are statistical only. Also shown in Figure 16 are Monte Carlo predictions. The results for the neutron production cross sections are summarized in Table 5, which includes both statistical and systematic errors for data. Monte Carlo predictions include results from DPMJET(p+p), FLUKA(p+A), and LAQGSM models. FLUKA cross sections were calculated in two ways: as a fraction of the inelastic cross section, and also using the normalized yield per target nucleus per incident proton. Both approaches gave results consistent with 1%. The DPMJET/FLUKA Monte Carlo gives cross sections in reasonable agreement with our measurements within our overall uncertainties that are dominated by the systematic errors. The systematic errors for data are ~15% while for the Monte Carlo, uncertainties in the FLUKA predictions are ~4% as estimated by varying the beam positions and widths. Comparisons of data with DPMJET/FLUKA/LAQGSM predictions are shown in Figure

16. Finally, the *A*-dependence of the cross section is shown in Figure 17 along with the behavior predicted by the Monte Carlos. The cross sections were fit to $A^{\alpha}$ where for data we found that $\alpha$ is $0.46 \pm 0.06$ for 58 GeV/c beam and $0.54 \pm 0.05$ for 120 GeV/c. A fit to the FLUKA cross sections gives $\alpha = 0.59$ at 58 GeV/c and 0.67 at 120 GeV/c. For LAQGSM, $\alpha = 0.40$ and 0.41 at 58 and 120 GeV/c. Thus, FLUKA predicts a steeper dependence on atomic weight than our measurements, while the LAQGSM model shows a flatter dependence. The neutron production cross sections predicted by LAQGSM for our geometry are typically ~60% of the measured ones.

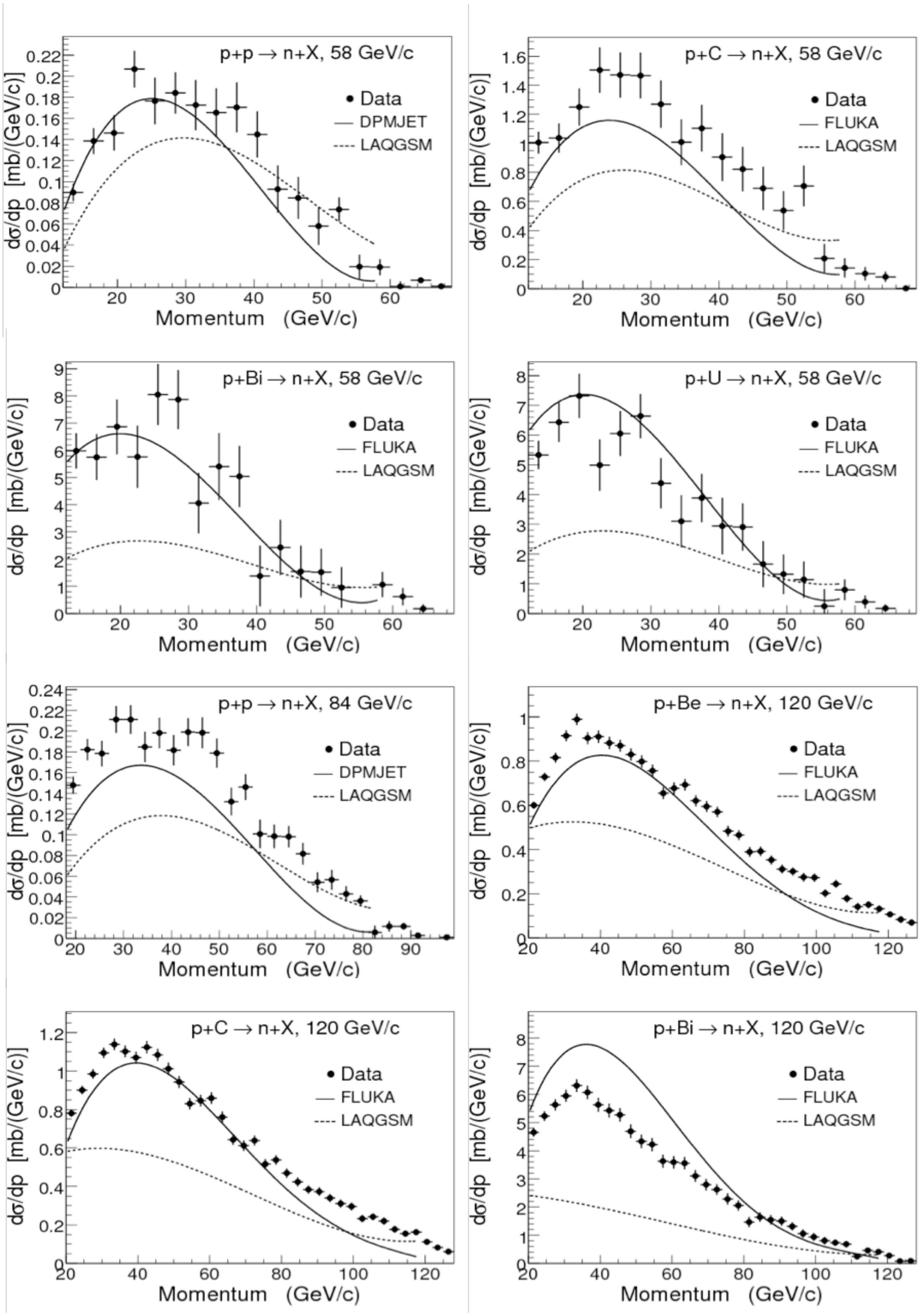

**FIGURE 16.** Measured cross sections from this experiment compared with neutron spectra generated by the DPMJET/FLUKA and LAQGSM Monte Carlos. Data and Monte Carlos are not corrected for geometric acceptance. The error bars are statistical only.

| Target, Beam Momentum | $\sigma_n$ (mb) Data ± Stat ± Syst | $\sigma_n$ (mb) DPMJET/FLUKA | $\sigma_n$ (mb) LAQGSM |
|---|---|---|---|
| H, 58 GeV | 5.8 ± 0.2 ±1.0 | 4.9 | 4.7 |
| C, 58 GeV | 45.9 ± 1.5 ± 7.2 | 33.3 | 27.5 |
| Bi, 58 GeV | 193.2 ± 10.4 ± 32.8 | 176.6 | 88.3 |
| U, 58 GeV | 178.8 ± 7.7 ± 31.3 | 198.1 | 91.5 |
| H, 84 GeV | 8.8 ± 0.2 ± 1.2 | 6.4 | 5.4 |
| Be, 120 GeV | 55.0 ± 0.4 ± 7.9 | 45.4 | 33.4 |
| C, 120 GeV | 64.5 ± 0.4 ± 8.7 | 56.1 | 36.6 |
| Bi, 120 GeV | 300.2 ± 3.1 ± 48.8 | 381.6 | 120.1 |

**TABLE 5.** Comparison of our measurements with Monte Carlo. Listed are cross sections for producing neutrons with energy greater than threshold and within an angular range of 20.4 mrad. These cross sections are not corrected for geometric acceptance.

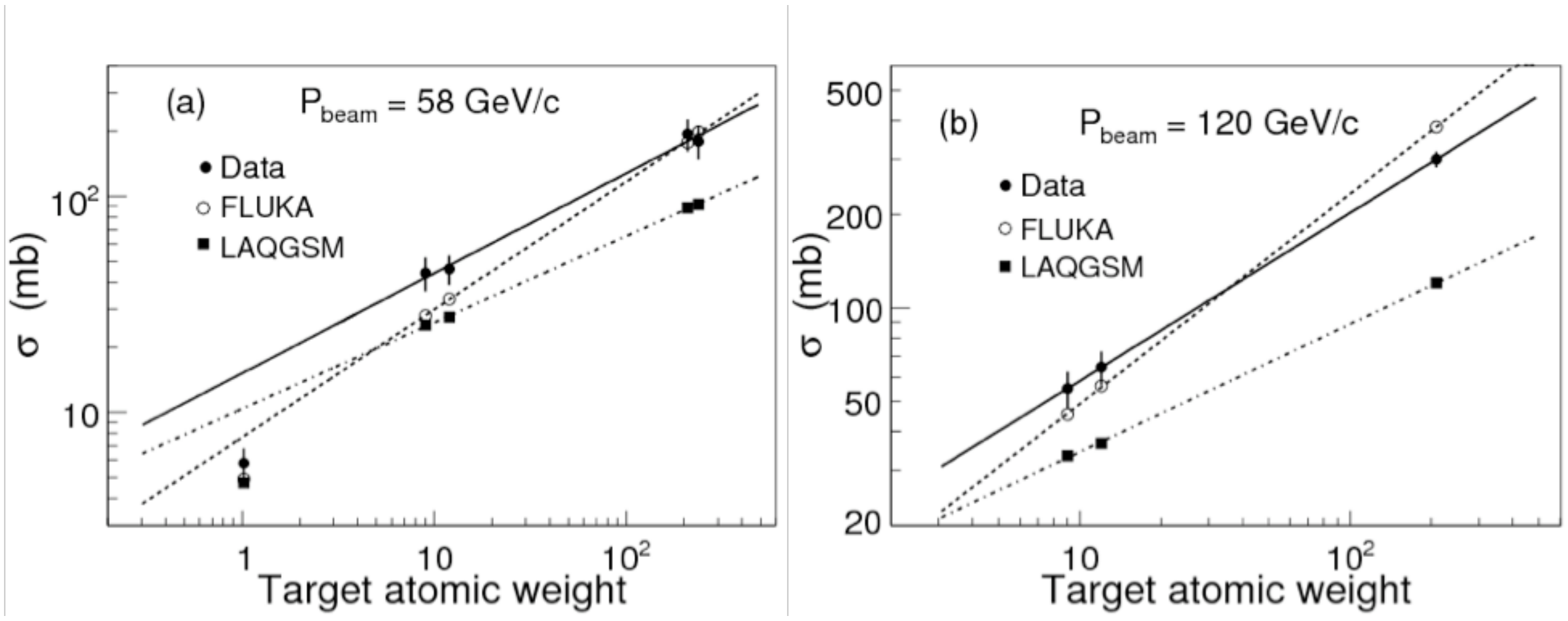

**FIGURE 17.** Comparison of the *A*-dependence of MIPP cross sections with those from Monte Carlos. The lines are fits to the data. The cross sections are for producing neutrons with momentum greater than the threshold and within an angular range of 20.4 mrad. The errors are combined statistical and systematic uncertainties. Note that the hydrogen data point is not included in the fit. The cross sections are not corrected for geometric acceptance.

One can also calculate the Lorentz-invariant cross section for producing neutrons as follows:

$$E\frac{d^3\sigma}{dp^3} = \frac{E}{p^2\Omega}\frac{d\sigma}{dp} \tag{2}$$

where $\Omega$ is the solid angle subtended by the calorimeter, $\approx 0.0013$ steradians, and $\frac{d\sigma}{dp}$ is calculated as described in Eq. (1) above. The Lorentz-invariant cross section distributions in $x_F$ are shown in Figure 18. $x_F$ was calculated as $x_F = (p_z / p_0)_{CM}$ where $p_z$ is the neutron's longitudinal momentum and $p_0$ is the incident beam momentum in the center of mass. In Figure 19 we compare the Lorentz-invariant cross sections for the hydrogen, carbon, and bismuth targets at different beam momenta. The plots indicate that the Lorentz-invariant inclusive cross sections show scaling for the hydrogen target but scaling has not been reached for the other targets.

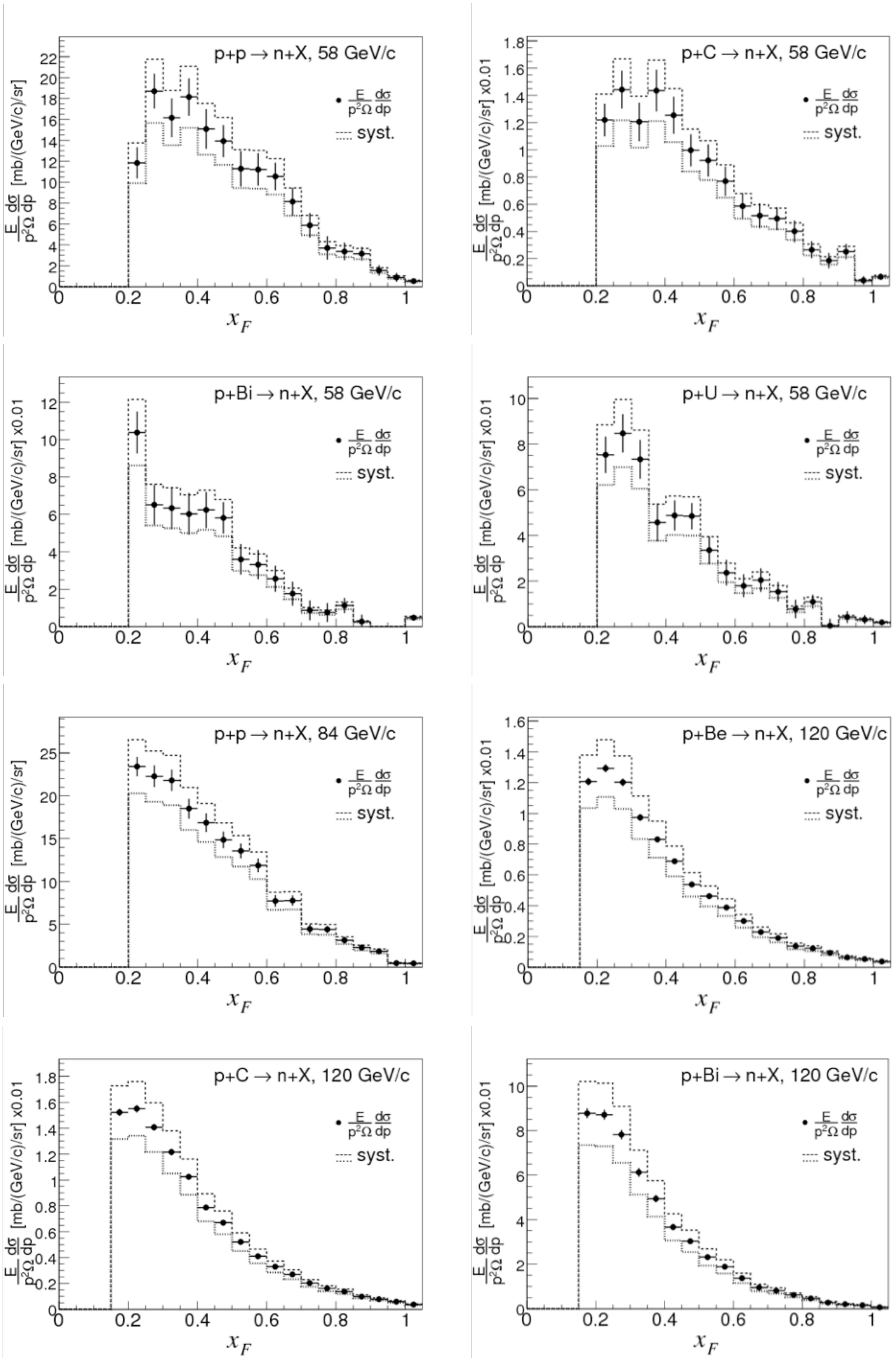

**FIGURE 18.** Lorentz-invariant cross-sections as a function of $x_F$ for producing neutrons with momentum greater than threshold within the solid angle of the calorimeter.

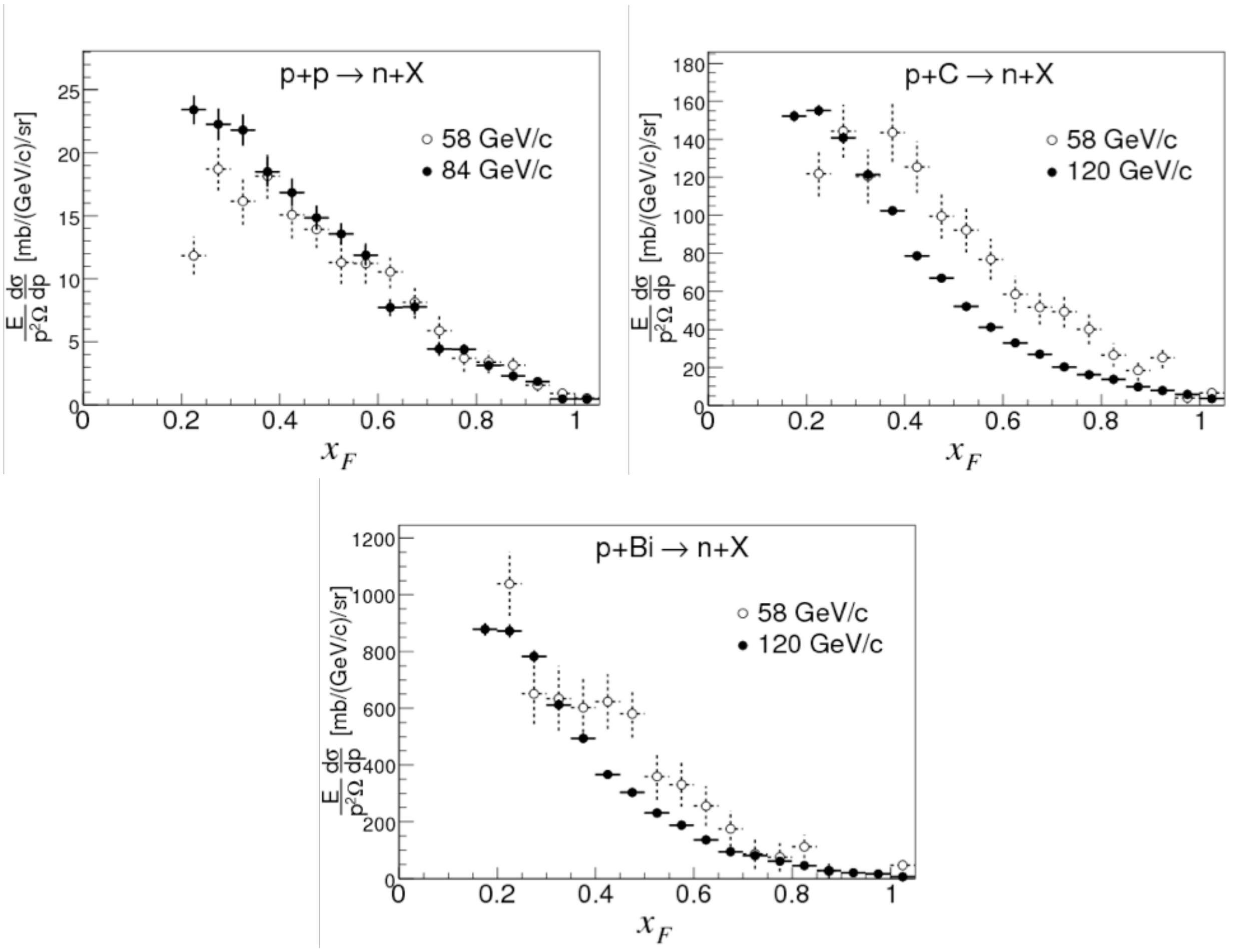

**FIGURE 19.** Lorentz-invariant cross sections as a function of $x_F$ for hydrogen, carbon and bismuth targets at different incident beam momenta. Errors are statistical only.

We also report the cross section for producing neutrons with energy greater than the threshold. The cross sections for neutrons within the fiducial volume of the calorimeter have already been shown in Fig. 16 above. These cross sections were then corrected for the geometric acceptance of the calorimeter. The acceptances were derived from FLUKA/DPMJET Monte Carlos as described earlier. These are summarized in Table 6 and the distributions in $x_F$ are shown in Figure 20.

| | $\sigma_n$ (mb) | Uncertainties (mb) stat,syst |
|---|---|---|
| H, 58 GeV | 9.6 | ± 0.3, ± 1.9 |
| C, 58 GeV | 82.6 | ± 2.7, ± 17.0 |
| Bi, 58 GeV | 468.3 | ± 25.2, ± 114.5 |
| U, 58 GeV | 437.1 | ± 18.9, ± 108.8 |
| H, 84 GeV | 11.1 | ± 0.2, ± 1.6 |
| Be, 120 GeV | 62.6 | ± 0.4, ± 9.3 |
| C, 120 GeV | 74.1 | ± 0.5, ± 10.4 |
| Bi, 120 GeV | 379.5 | ± 3.9, ± 66.1 |

**TABLE 6.** Neutron production cross sections per nucleus and their uncertainties. The distinction between the numbers in this table and those in Table 5 is that here we report the production cross sections after correcting for the geometric acceptance of our detector.

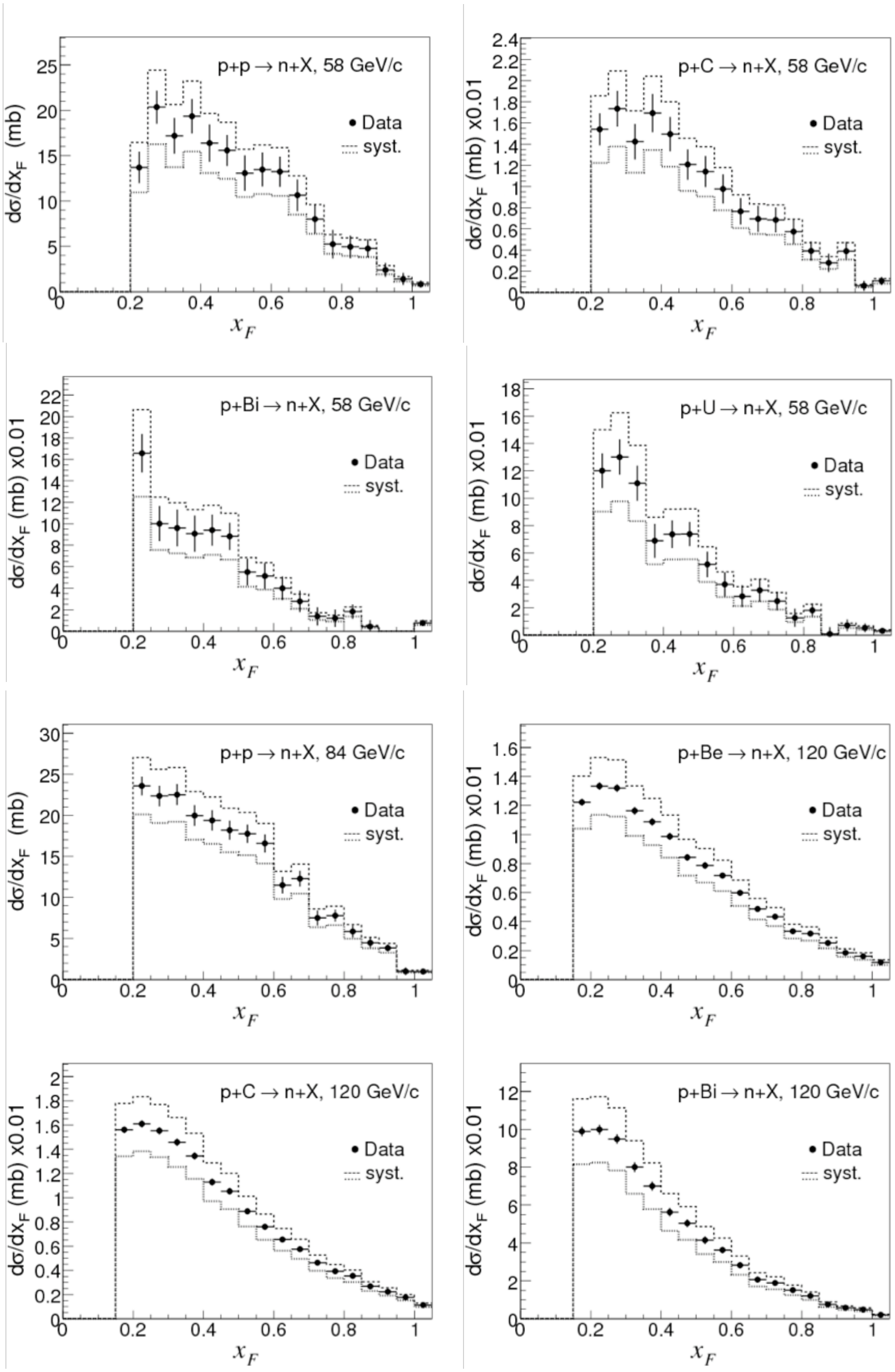

**FIGURE 20.** Total inclusive cross sections as a function of $x_F$ for producing neutrons with momentum greater than the threshold with estimated systematic uncertainties. The difference between these plots and the ones in Figure 16 is that here we have corrected for the geometric acceptance of the calorimeter.

## VIII. COMPARISON WITH OTHER MEASUREMENTS

There are very little data on neutron production. For some of the published data, the measurements are at discrete production angles making it impossible to directly compare them with our results [20,21,22]. We have compared our measurements with the results from the NA49 experiment that measured neutron production from p+p collisions at 158 GeV/c [23]. Figure 21 shows the comparison of the neutron density, $dn / dx_F$ , with our hydrogen data at 58 GeV/c and 84 GeV/c beam momenta. $dn / dx_F$ was calculated as:

$$\frac{dn}{dx_F} = \frac{1}{\sigma_{inelastic}} \frac{d\sigma}{dx_F} \tag{3}$$

Both measurements have been corrected for acceptance and efficiencies. There is reasonable agreement at $x_F > 0.6$, but for lower $x_F$ the NA49 data at 158 GeV/c are lower than both our 58 and 84 GeV/c data by a factor ~2 . This discrepancy is not understood.

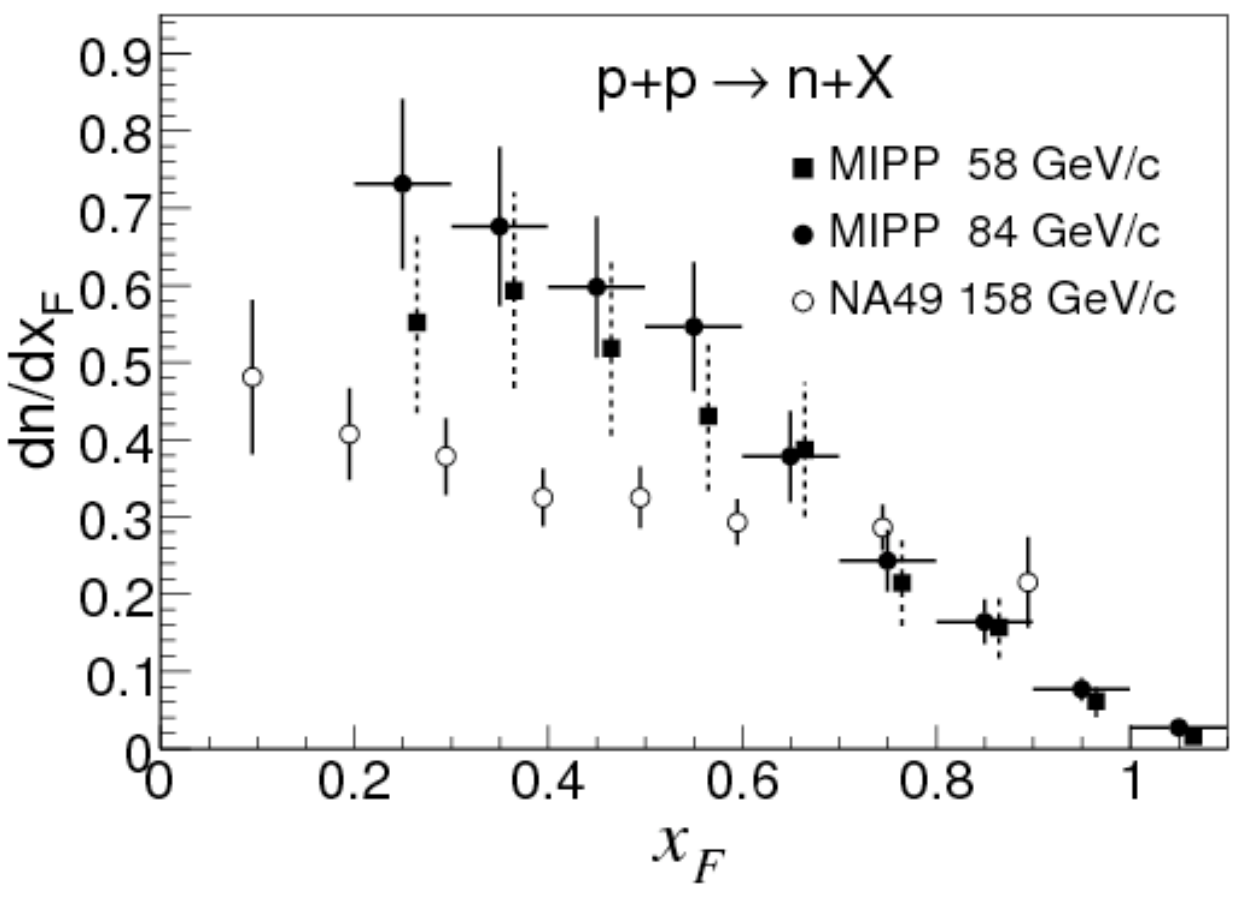

**FIGURE 21.** Comparison of our results (solid points) for 58 GeV/c and 84 GeV/c incident protons with those from NA49 (open circles). The errors in our data are combined statistical and systematic uncertainties.

## IX. CONCLUSIONS

We have measured neutron production cross sections from 58 GeV/c to 120 GeV/c protons incident on hydrogen, beryllium, carbon, bismuth and uranium targets. We have also reported Lorentz-invariant neutron production cross sections. The neutron cross sections integrated over $p_t$ show reasonable agreement with the DPMJET/FLUKA simulation. The LAQGSM Monte Carlo predicts cross sections that are lower than our measurements and FLUKA. This discrepancy is under study [24]. The shape of our $p_T$ integrated neutron yield agrees with the results from NA49 [23] for $x_F$>0.6. At lower $x_F$ the NA49 spectrum is suppressed and shows a slower rise with decreasing $x_F$ in comparison with ours. We have also found that for atomic weight $A$ > 1 the total neutron production cross sections vary as $A^{\alpha}$ where $\alpha$ is $0.54 \pm 0.05$ for 120 GeV/c beam and $0.46 \pm 0.06$ for 58 GeV/c beam.

## ACKNOWLEDGMENTS

The efforts of the Fermilab staff are gratefully acknowledged. We are grateful to N. V. Mokhov, S.I. Striganov, and K. K. Gudima for providing the LAQGSM simulations. This research was sponsored by the National Nuclear Security Administration under the Stewardship Science Academic Alliances program through DOE Research Grant DE-FG52-2006NA26182 and the U. S. Department of Energy.